\makeatletter
\declare@file@substitution{revtex4-1.cls}{revtex4-2.cls}
\makeatother
\documentclass[twocolumn, times, tighten,twocolappendix]{aastex63}
\usepackage{graphicx,multirow,hyperref,url,color,xspace}
\usepackage{amsmath,amssymb}
\DeclareMathAlphabet      {\mathbfit}{OML}{cmm}{b}{it}

\newcommand{\msun}{{\rm M}_{\sun}}
\newcommand{\rsun}{{\rm R}_{\sun}}

\newcommand{\appropto}{\mathrel{\vcenter{
 \offinterlineskip\halign{\hfil$##$\cr
 \propto\cr\noalign{\kern2pt}\sim\cr\noalign{\kern-2pt}}}}}

\defcitealias{Miller-Jones21}{MJ21}
\newcommand{\mj}{{\citetalias{Miller-Jones21}}\xspace}

\begin{document}

\title{Evidence for a black hole spin--orbit misalignment in the X-ray binary Cyg X-1}
\shorttitle{Black hole spin--orbit misalignment in Cyg X-1}

\author[0000-0002-0333-2452]{Andrzej A. Zdziarski}
\affiliation{Nicolaus Copernicus Astronomical Center, Polish Academy of Sciences, Bartycka 18, PL-00-716 Warszawa, Poland; \href{mailto:aaz@camk.edu.pl}{aaz@camk.edu.pl}}

\author[0000-0002-5767-7253]{Alexandra Veledina}
\affiliation{Department of Physics and Astronomy, FI-20014 University of Turku, Finland}
\affiliation{Nordita, KTH Royal Institute of Technology and Stockholm University, Hannes Alfv{\' e}ns v{\" a}g 12, SE-10691 Stockholm, Sweden}

\author[0000-0001-7606-5925]{Micha{\l} Szanecki}
\affiliation{Faculty of Physics and Applied Informatics, {\L}{\'o}d{\'z} University, Pomorska 149/153, PL-90-236 {\L}{\'o}d{\'z}, Poland}

\author{David A. Green}
\affiliation{Cavendish Laboratory, 19 J. J. Thomson Ave., Cambridge, CB3 0HE, UK}

\author{Joe S. Bright}
\affiliation{Astrophysics, Department of Physics, University of Oxford, Keble
Road, Oxford OX1 3RH, UK}

\author{David R. A. Williams}
\affiliation{Jodrell Bank Centre for Astrophysics, School of Physics and
Astronomy, The University of Manchester, Manchester, M13 9PL, UK}

\shortauthors{Zdziarski et al.}

\begin{abstract}
Recently, the accretion geometry of the black-hole X-ray binary Cyg X-1 was probed with the X-ray polarization. The position angle of the X-ray emitting flow was found to be aligned with the position angle of the radio jet in the plane of the sky. At the same time, the observed high polarization degree could be obtained only for a high inclination of the X-ray emitting flow, indicating a misalignment between the binary axis and the black hole spin. The jet, in turn, is believed to be directed by the spin axis, hence similar misalignment is expected between the jet and binary axes. We test this hypothesis using very long (up to about 26 years) multi-band radio observations. We find the misalignment of $20\degr$--$30\degr$. However, on the contrary to the earlier expectations, the jet and binary viewing angles are found to be similar, while the misalignment is seen between position angles of the jet and the binary axis on the plane of the sky. Furthermore, the presence of the misalignment questions our understanding of the evolution of this binary system.
\end{abstract}

\section{Introduction}
\label{intro}

The archetypical high-mass X-ray binary Cyg X-1, discovered as an X-ray source in 1964 \citep{Bowyer65}, is probably the best studied microquasar to date. We have accurate determination of the binary parameters, including the orbital period of $P=5.599829$\,d \citep{Brocksopp99}, and other parameters (given here as the median values with 68\% uncertainties), including the binary inclination, $i=27\fdg5_{-0.6}^{+0.8}$, the mass of the black hole (BH), $M_{\rm BH}=21.2\pm 2.2 M_{\odot}$, the donor mass, $M_*=40.6^{+7.7}_{-7.1} M_{\odot}$, and the distance to the source, $D=2.22^{+0.18}_{-0.17}$\,kpc (\citealt{Miller-Jones21}, hereafter \mj). Moreover, the spin parameter of the BH has been measured as $a_*\gtrsim 0.5$ \citep{Kawano17} and $\lesssim 1$ (\mj). This BH spin could not be acquired during accretion given the short life time of the system, which implies it was acquired before the BH formation. Moreover, the low proper motion of Cyg X-1 with respect to its likely parent association Cyg OB3 of $10.7\pm 2.7$\,km\,s$^{-1}$ \citep{Rao20} indicates the BH was formed with a low natal kick. As then estimated by \mj, the BH spin axis appears to be inclined by at most $10\degr$ from the axis of the binary.

With the launch of the {\it Imaging X-ray Polarimetry Explorer} ({\it IXPE}; \citealt{Weisskopf2022}), the issue of a possible misalignment was revived. The polarimetric measurements of Cyg X-1 by {\it IXPE\/} implied that the X-rays are produced in a hot gas that is flattened in the direction orthogonal to the resolved relativistic jet, observed in the system \citep{Krawczynski22}. This configuration corresponds either to the truncated disc geometry \citep{PKR97,Esin97} or to the slab corona geometry \citep{HM91}, whose axis is well aligned in the plane of the sky with the position angle of the jet. The observed high, $\approx$4\%, polarization degree cannot be explained if the inclination of this hot inner region is equal to the orbital inclination; it can be achieved instead if the inclination is higher, $\gtrsim\! 45\degr$. This discrepancy may indicate a misalignment of the BH spin with respect to the binary axis, implying a geometry where the outer parts of the disc are aligned with the orbital axis and the inner accretion flow is aligned with the BH spin \citep{Bardeen75}.

On the other hand, the position angle of the binary, if assumed equal to the optical polarization angle \citep{Krawczynski22}, shows good agreement with the X-ray polarization angle and with the position angle of the jet. This indicates that the orbital axis and the BH spin coincide in the plane of the sky. The misalignment would thus be evident only along the line of sight direction.

Since jets are launched along the spin axis of the BH \citep{BZ77, McKinney13}, a BH spin--orbit misalignment should be visible as a jet--orbit one. In the case of jets launched from discs \citep{BP82}, this will be the case if the inner disc is aligned with the BH spin \citep{Bardeen75}. The arguments above suggest that the jet axis should be inclined at more than $45\degr$ with respect to the observer, i.e., misaligned from the orbital plane by $\gtrsim\! 20\degr$, but in the plane of the sky the jet direction should coincide with the orbital angular momentum vector. Using the long-term radio light curves of Cyg X-1, we show that this picture is reversed. Namely, we find here a compelling evidence for a jet--orbit misalignment in Cyg X-1, which manifests itself on the plane of the sky, while the inclination angles of the binary orbit and the jet coincide within a few degrees.

\section{The data}
\label{data}

The radio jet in Cyg X-1 is relatively steady and compact in the hard and hard-intermediate X-ray spectral states \citep{DGK07} of the source, and is much weaker during the soft states (e.g., \citealt{ZSP20}), as is typical for accreting BH binaries \citep{FBG04}. 

The radio emission is stable on average, but shows modulations at the orbital period. Study of the initial 20-month radio light-curve of the Ryle Telescope \citep{Jones91} revealed strong periodic modulation of emission at 15~GHz \citep{Pooley99} with a fractional semi-amplitude of $\approx$0.17. The study found a tentative evidence for a lag of the minimum radio flux with respect to the superior conjunction (the phase of the orbit when the compact object is located behind the star) by $\approx$0.12 of the orbital period, and, using the Green Bank Interferometer data at 8.3 and 2.25 GHz, suggested this lag to be increasing with the decreasing frequency. A promising interpretation for the lag is the misalignment of the jet axis from that of the binary \citep{Malzac09}. The quality of the data, available at that time, however, did not allow for a detailed modelling. By now, the amount of the available 15 GHz data from the Ryle Telescope and its successor, the Arcminute Microkelvin Imager Large Array (AMI-LA) \citep{Hickish18}, has increased many-fold. The unprecedentedly long duration of the resulting light curve enables detailed modelling of the orbital modulation, including the accurate determination of the lag. 

In our study, we use the radio light curves at 15\,GHz from the Ryle Telescope and the AMI-LA (jointly covering MJD 50226--59575), and at 8.3 and 2.25 GHz, from Green Bank Interferometer (covering MJD 50409--51823) in the hard and hard-intermediate spectral states only. The hard and hard intermediate state intervals are defined based on the X-ray hardness as in \citet{ZSP20} and are given in Table~\ref{dates}. The average fluxes in those states at 15, 8.3 and 2.25\,GHz are $\langle F_\nu\rangle=12.5$, 15.0 and 14.3\,mJy, respectively. 

\setlength{\tabcolsep}{1.pt}
\begin{table}
\centering  
\caption{The adopted intervals (in MJD) of the occurrences of the hard and intermediate states.}\vskip -0.5cm
\begin{tabular}{cccccccccccccc}
\hline
Start & End &\hbox to 0.05cm{\hfill}& Start & End &\hbox to 0.05cm{\hfill}& Start & End &\hbox to 0.05cm{\hfill}& Start & End &\hbox to 0.05cm{\hfill}& Start & End\\
\hline
50085 &50222 &&52853 &53003 &&55895 &55940 &&57105 &57265 &&58631 &58792\\
50308 &51845 &&53025 &53265 &&56035 &56087 &&57332 &57970 &&59378 &59391\\
51858 &52167 &&53292 &53368 &&56722 &56748 &&58112 &58210 &&59420 &59860\\
52205 &52237 &&53385 &55387 &&56760 &56845 &&58387 &58416\\
52545 &52801 &&55674 &55790 &&57012 &57045 &&58482 &58585\\
\hline
\end{tabular}
\label{dates}
\end{table} 

We fold and average the light curves over the ephemeris of
\begin{equation}
t_{\rm sup}=t_0 + P m,\, t_0=50077.973, \, P=5.599829\,{\rm d},
\label{eph}
\end{equation}
where $t_{\rm sup}$ is the time of a superior conjunction (the black hole furthest from the observer) in MJD, and $m$ is an integer. This $t_0$ is below the start times of our light curves, and it fully agrees with the previously given $t_0$ \citep{Gies08}. 

\section{The model}
\label{model}

We use the median values of $M_{\rm BH}$, $M_*$, $D$, and $i$ as given in Section \ref{intro}. The donor radius is $R_*\approx 22\rsun$ and its effective temperature is $T_*\approx 3.1\times 10^4$\,K (\mj). The masses and the orbital period imply the semi-major axis of $a\approx 53\rsun$.

The modulation is due to free--free absorption of the radio emission by the stellar wind from the supergiant donor \citep{Walborn73}, with the path toward the observer through the wind being orbital phase-dependent. The amplitudes of the modulation can be explained if the radio emission originates in the jet at heights comparable to the stellar separation \citep{SZ07, Zdziarski12}. The strongest absorption corresponds to the highest column density of the wind along the path to the observer. For a jet perpendicular to the plane of the sky, this would correspond to the superior conjunction. However, when the jet is significantly inclined and its approaching part lags behind the binary axis (as projected on the sky), the highest column density occurs at an orbital phase {\it after} the superior conjunction. In this Letter, we show that the shapes of the orbital modulation can be well fitted if the jet is misaligned with respect to the binary axis. 

\begin{figure}
\centerline{\includegraphics[width=\columnwidth]{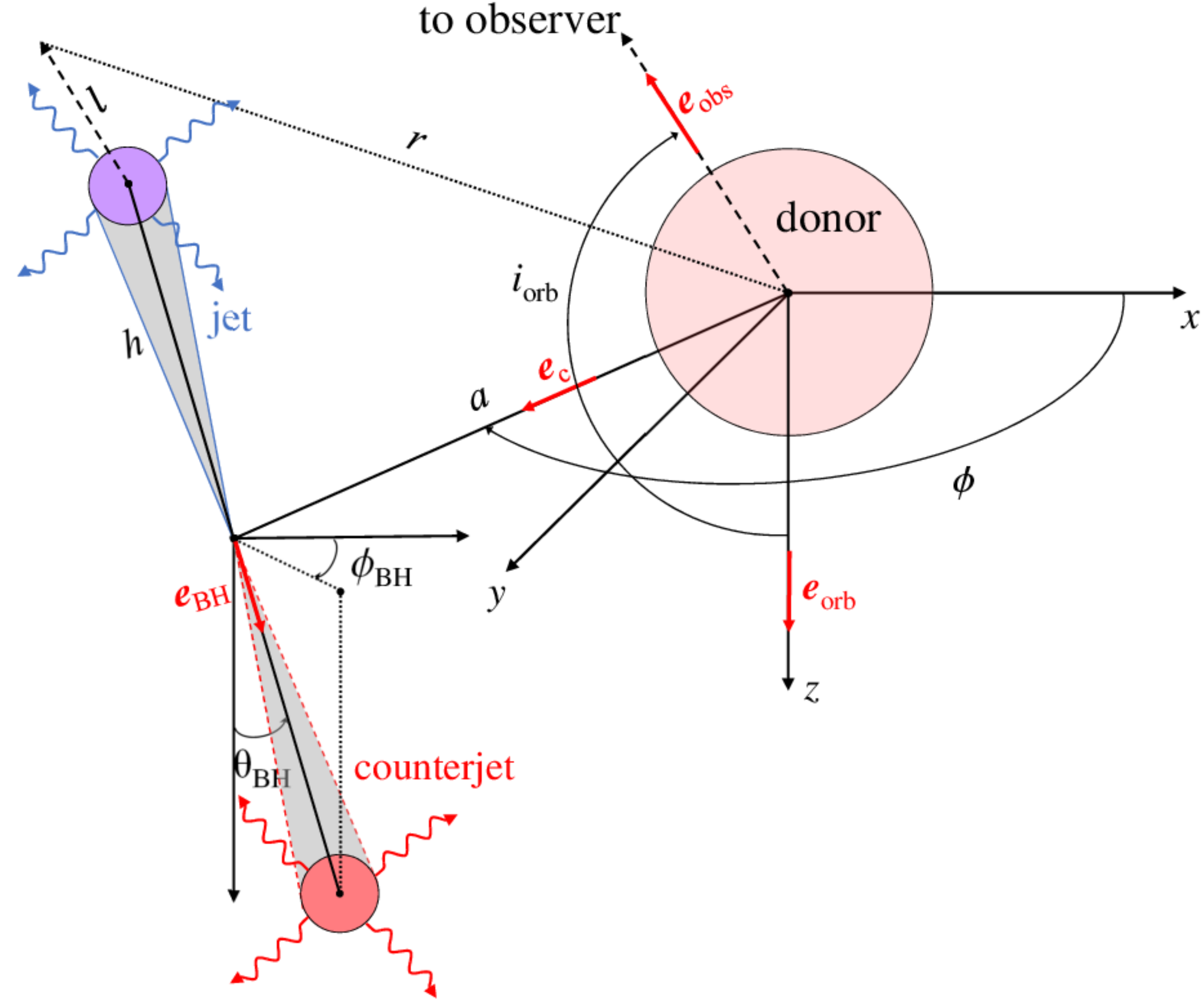}}
\caption{The geometry of the binary and the jets. The axes $x$ and $y$ are in the binary plane, and $+z$ gives the direction along the binary vector, ${\mathbfit e}_{\rm orb}$ (away from the observer, given the observed clockwise rotation). The observer is at an angle, $i_{\rm orb}$, with respect to ${\mathbfit e}_{\rm orb}$, and $\phi$ is the orbital phase; $\phi=0$ and $\pi$ correspond to the superior and inferior conjunction, respectively. The binary rotation follows the increasing $\phi$. The shown configuration is close to the latter. Then, $\theta_{\rm BH}$ is the inclination of the BH spin vector, ${\mathbfit e}_{\rm BH}$, and the counterjet with respect to ${\mathbfit e}_{\rm orb}$, $\phi_{\rm BH}$ is its azimuthal angle, and $h$ is the distance of the radio source from the BH center. The counterjet emission is neglected in our model. The distance from the radio source measured along the direction toward the observer is $l$. The distances of the point at $l$ measured from the centers of the BH and the donor are $s$ (not shown for clarity) and $r$, respectively.
} \label{geometry}
\end{figure}

We specify the coordinate system and the geometry in Figure~\ref{geometry}. For simplicity, we assume the orbit to be circular, since the eccentricity is only $\approx\!0.019\pm 0.003$ (\mj). Movement of the Cyg X-1 binary components in the orbit is clockwise on the sky (\mj). This implies that the orbital spin vector (along the $+z$ axis) points away from us, and the inclination of that direction is $i_{\rm orb}=180\degr - i$, where $i$ is the binary inclination measured using spectroscopic and photometric data (which are not sensitive to the orientation of the orbit). The estimates of its BH spin of $a_*>0$ imply that rotation of the accretion flow is prograde. Thus, the direction of the BH spin vector, and consequently, of the jet spin, also points away from the observer, and along the counterjet. The standard convention that the superior conjunction corresponds to the orbital phase of $\phi=0$ requires that the projection of the direction toward the observer onto the binary plane is in the $-x$ direction. The inclination of the BH spin with respect to the binary axis is $\theta_{\rm BH}$, which represents the misalignment of the BH spin vector (and the counterjet) with respect to that of the binary, as well as the misalignment of the jet with respect to the $-z$ direction. The angle of the projection of the BH spin vector onto the binary plane with respect to the $+x$-axis is the azimuthal angle $\phi_{\rm BH}$.

Following the above, the unit vectors pointing towards the observer, from the stellar center to the compact object, and along the jet are,
\begin{align}
&{\mathbfit e}_{\rm obs}=(-\sin i_{\rm orb}, 0, \cos i_{\rm orb}),\quad {\mathbfit e}_{\rm c}=(\cos\phi,\sin\phi, 0),\nonumber\\
&{\mathbfit e}_{\rm BH}=(\sin\theta_{\rm BH}\cos\phi_{\rm BH}, \sin\theta_{\rm BH}\sin\phi_{\rm BH} , \cos\theta_{\rm BH}),\label{vectors}
\end{align}
respectively, where $i_{\rm orb}=180\degr - i$. Hereafter, we neglect the emission of the counterjet, since \citet{Stirling01} gives the jet/counterjet ratio of $\approx$50. We make a simplifying assumption that the modulated emission at a given frequency originates, on average, at a single distance, $h$, from the BH. We calculate the optical depth starting at the emission point on the jet at the distance $h$ from the BH along the direction toward the observer. The vector connecting the donor center with a point along the photon trajectory at the distance $l$ from the emission point is $a {\mathbfit e}_{\rm c}- h {\mathbfit e}_{\rm BH}+ l {\mathbfit e}_{\rm obs}$, and the vector connecting the center of the BH with that point at $l$ is $-h {\mathbfit e}_{\rm BH}+ l {\mathbfit e}_{\rm obs}$. The lengths of these vectors give the distances of that point from the donor and BH centers, and their squares are given by
\begin{align}
&r^2 = (l \cos i_{\rm orb} -h \cos\theta_{\rm BH})^2+(l \sin i_{\rm orb}-a \cos\phi+\nonumber \\
&h \sin\theta_{\rm BH}\cos\phi_{\rm BH})^2+(h \sin\theta_{\rm BH}\sin\phi_{\rm BH}-a \sin\phi)^2, \label{r2}\\
&s^2=h^2+l^2-2 h l \cos i_{\rm orb} \cos\theta_{\rm BH}+\nonumber \\
&2 h l \sin i_{\rm orb} \sin\theta_{\rm BH}\cos\phi_{\rm BH},\label{x2}
\end{align}
respectively. The cosine of the viewing angle of the BH spin vector, $i_{\rm BH}$, is given by ${\mathbfit e}_{\rm BH}\cdot {\mathbfit e}_{\rm obs}$, and that of the jet, by $-{\mathbfit e}_{\rm BH}\cdot {\mathbfit e}_{\rm obs}$. The viewing angle of the jet and of the BH spin are then
\begin{align}
&i_{\rm jet}=\arccos (-\cos i_{\rm orb} \cos\theta_{\rm BH} + \cos\phi_{\rm BH} \sin i_{\rm orb} \sin\theta_{\rm BH}),\nonumber\\
&i_{\rm BH}=180\degr - i_{\rm jet},\label{ibh}
\end{align}
respectively.

We calculate the expected position angle of the projection of the orbital axis on the sky, $\lambda_{\rm orb}$, with respect to the projection of the BH spin, $\lambda_{\rm BH}$. The difference between these two angles \citep{Poutanen22} for clockwise rotation on the sky is
\begin{align}
&\Delta\lambda\equiv \lambda_{\rm BH}-\lambda_{\rm orb}= \arccos\frac{\cos\theta_{\rm BH} - \cos i_{\rm BH}\cos i_{\rm orb}}{\sin i_{\rm BH}\sin i_{\rm orb}}=\nonumber\\
&\arccos\frac{\cos i_{\rm orb}\cos \phi_{\rm BH} \sin \theta_{\rm BH}+\sin i_{\rm orb} \cos
\theta_{\rm BH}}{\sqrt{1-\left(\cos i_{\rm orb} \cos \theta_{\rm BH}-\sin i_{\rm orb}\cos \phi_{\rm BH} \sin \theta_{\rm BH}\right)^2}}\!,
\label{delta}
\end{align}
and the observed jet position angle is $\lambda_{\rm jet}= \lambda_{\rm BH}\pm 180\degr$. Thus, $\lambda_{\rm orb}=\lambda_{\rm jet}-\Delta\lambda\pm 180\degr$.

Next, we calculate the free--free absorption from a point along the jet at the distance $h$ from the BH center toward the observer, i.e., along the line denoted $l$ in Figure~\ref{geometry}. We assume an isotropic stellar wind with the standard velocity profile \citep{Lamers87},
\begin{equation}
v(r) \simeq v_{\infty}\left(1-\frac{R_*}{r}\right)^{\beta},
\label{velocity}
\end{equation}
where $v_\infty$ is the terminal wind velocity and the exponent $\beta$ determines the acceleration rate. The electron density, $n$, follows then from the continuity equation at a given mass loss rate, $\dot M$. For simplicity, the presence of ions heavier than hydrogen is neglected. We use $v_\infty =1.6\times 10^8$\,cm\,s$^{-1}$, $\beta=1$ \citep{Gies86b}, and $\dot M=-2.6\times 10^{-6}\msun$\,yr$^{-1}$ in the hard spectral state \citep{Gies03}. However, given the likely decrease of the wind density in the polar region (\citealt{Gies08}; crossed by the line of sight of the radio photons), we scale that $\dot M$ by a factor $f\leq 1$. 

The free--free absorption coefficient is approximately \citep{RL79},
\begin{equation}
\alpha_{\rm ff} \approx 0.12\! \left(\frac{T}{1\,{\rm K}}\right)^{-3/2}\!\! \left(\frac{n}{1\,{\rm cm}^{-3}}\right)^2\! \left(\frac{\nu}{1\,{\rm GHz}}\right)^{-2}\! {\rm cm}^{-1}.
\label{alpha_ff}
\end{equation}
The phase-dependent optical depth to free--free absorption is
\begin{align}
&\tau(\phi)=\tau_0 \left(\frac{\nu}{15\,{\rm GHz}}\right)^{-2}\times \nonumber\\
&\int_0^\infty \left[r(l)\over a\right]^{-4} \left[1-{R_* \over r(l)}\right]^{-2\beta} \!\!\left\{T[r(l),s(l)]\over T_0\right\}^{-3/2}\!\!
\frac{{\rm d}l}{a},
\label{tau_v}
\end{align}
where $\tau_0$ is a reference optical depth defined at 15\,GHz, the density at $r=a$ under the assumption of $v=v_\infty$, the distance $a$, and at a reference temperature, $T_0$,
\begin{align}
&\tau_0\approx 26.3 \left( -f \dot{M}\over 2.6\times 10^{-6}\,\msun\,{\rm yr}^{-1}\right)^2 \left(M_*+M_{\rm BH}\over 62\msun\right)^{-1}\times \nonumber\\
&\left(v_\infty\over 1.6\times 10^8\,{\rm cm\ s}^{-1}\right)^{-2} \left(T_0\over 10^6\,{\rm K}\right)^{-3/2}. 
\label{tauref}
\end{align}
The observed flux is
\begin{equation}
F(\phi)=F_{\rm intr}\exp[-\tau(\phi)],
\label{F_phi}
\end{equation}
where $F_{\rm intr}$ is the flux before the absorption. We calculate $\tau(\phi)$ numerically using $T(r,s)$ from the solution of the energy balance equation including Compton and photoionization heating and Compton, recombination and line cooling \citep{Zdziarski12}. The ionizing X-ray luminosity is estimated as $L_{\rm ion}\approx 2\times 10^{37}$\,erg\,s$^{-1}$.  The radiative heating and cooling of the wind at distance $l$ from the emission point is by both the emission of the donor and by the X-rays. However, we do not include the adiabatic cooling. Such cooling would lead to a strong temperature decrease \citep{Zdziarski12} at $r\gtrsim a$, and the associated strong increase of the absorption coefficient. Our neglect of that cooling is motivated by the wind expansion being, to a good approximation, in vacuum, with the wind density at radii of interest (e.g., $n\sim 10^{10}$\,cm$^{-3}$ at $r=a$) orders of magnitude higher than the density of the interstellar medium, which is also swept away by the wind. Thus, the wind does not perform a $p {\rm d}V$ work on the surrounding medium. The expansion can lead to particle distribution to be anisotropic and different from a Maxwellian, which minor effect we neglect. 

We consider then the spatial distribution of the radio emission along the jet. Resolved radio maps \citep{Stirling01, Rushton10} imply that a fraction $\approx$0.3--0.5 of the 8\,GHz emission is emitted at distances $\gtrsim\! 2\times 10^{14}$\,cm, which is $\gtrsim\! 50 a$. A similar constraint follows from the measured long radio lag with respect to X-rays \citep{Tetarenko19}. However, if most of the radio flux originated from such distances, orbital modulation would have been negligible. 

On the other hand, we can calculate the photospheric distance from jet models. The hard-state 2--220\,GHz spectrum of Cyg X-1 \citep{Fender00} is approximately flat with $\alpha\approx 0$ (where the flux density $F_\nu \propto \nu^\alpha$). The simplest, and widely adopted, model of such a spectrum is partially self-absorbed synchrotron emission with the distributions of both nonthermal electrons and the magnetic energy flux maintained along the jet \citep{BK79}. The location of the bulk of the emission at a given frequency in this model approximately corresponds to unit optical depth to synchrotron self-absorption. For that, we use a previous formulation \citep{Zdziarski22a}. For the parameters of Cyg X-1, that distance is given by
\begin{equation}
\frac{h_\nu}{a} \approx 2.8 \frac{15{\rm GHz}}{\nu} \!\left(\frac{\sin 27\fdg5}{\sin i_{\rm jet}}\right)^{\frac{5+p}{13+2 p}}\!\!\!
\left(\frac{\langle F_\nu\rangle}{10{\rm mJy}}\right)^{\frac{6+p}{13+2 p}}\!\!\!
\left(\frac{1\degr}{\Theta}\right)^{\frac{7+p}{13+2 p}}\!\!\!,\label{hnu}
\end{equation}
where $2.8 a\approx 1.0\times 10^{13}{\rm cm}$, $\Theta$ is the jet opening angle, which has been estimated as $\approx$0\fdg4--$1\fdg8$ \citep{Tetarenko19}, and $p$ is the steady-state power-law index of the relativistic electrons, which is consistent with $\approx 2.5$--3.5 \citep{Zdziarski14c}. The dependencies of $h_\nu$ on the ratio of the gas-to-magnetic energy densities and on the minimum and maximum energies of the relativistic electrons are very weak. The range of values implied by Equation (\ref{hnu}), $h_\nu\approx (1.6$--$5.4)a$, is $\ll 50 a$, and consistent with the observed strong orbital modulations. 

The above considerations imply a relatively complex radio emission profile, which we approximate as two separate regions of the radio emission. Namely, we split the folded and averaged radio light curves, $F_\nu(\phi)$, into the modulated, $F_\nu^{\rm mod}$, and unmodulated parts,
\begin{align}
&F_\nu(\phi)= F_\nu^{\rm mod}(\phi) +b \langle F_\nu\rangle,\quad 
\Delta F_\nu^{\rm mod}(\phi)=\Delta F_\nu(\phi),\nonumber\\ &\frac{\Delta F_\nu^{\rm mod}(\phi)}{F_\nu^{\rm mod}(\phi)}\approx \frac{\Delta F_\nu(\phi)}{(1-b) F_\nu(\phi)},
\label{split}
\end{align}
where $\Delta F_\nu(\phi)$ is the uncertainty of $F_\nu(\phi)$, respectively, and $b$ is the unmodulated fraction. We use $b=0.5$ as a likely value. The unmodulated part accounts for the remote emission. However, we also perform the calculations at $b=0$. We find the two sets of the fitted parameters are very similar (see Table~\ref{fits_MCMC} below), which shows our results are not sensitive to the assumed value of $b$.

\section{Results}
\label{results}

\setlength{\tabcolsep}{2.pt}
\begin{table*}
\caption{Fit results based on the MCMC method. \label{fits_MCMC}
}
\centering\begin{tabular}{ccccccccccccccc}
\hline
$b$ & $h_{15}/a$ & $f$ & $i(\degr)$ & $\theta_{\rm BH}(\degr)$ & $\phi_{\rm BH}(\degr)$ & $\frac{\langle F_{\rm intr,15}\rangle}{\langle F_{15}\rangle}$ & $\frac{h_{8}}{h_{15}}$ & $\frac{\langle F_{\rm intr,8}\rangle}{\langle F_{8}\rangle}$ & $\frac{h_{2}}{h_{15}}$ & $\frac{\langle F_{\rm intr,2}\rangle}{\langle F_{2}\rangle}$ & $\frac{\chi^2}{{\rm d.o.f.}}$ & $i_{\rm jet}(\degr)$ & $\Delta\lambda(\degr)$ \\
\hline
0f & $2.9_{-1.8}^{+1.4}$ & $0.8_{-0.5}^{+0.2}$ & 27.5f &$23_{-8}^{+13}$ &$76_{-8}^{+16}$ &$1.24_{-0.06}^{+0.11}$ &$1.7_{-0.2}^{+0.5}$ &$1.15_{-0.07}^{+0.09}$ &$9.3_{-5.2}^{+9.5}$ &$1.02_{-0.02}^{+0.12}$ & 36/35 & $31_{-2}^{+6}$ & $47_{-17}^{+24}$
\vspace {0.1cm}\\
{\bf 0.5}f & $2.4_{-1.5}^{+1.1}$ & $0.8_{-0.4}^{+0.2}$ & 27.5f &$22_{-6}^{+11}$ &$76_{-8}^{+16}$ &$1.49_{-0.12}^{+0.15}$ &$1.7_{-0.2}^{+0.6}$ &$1.29_{-0.14}^{+0.17}$ &$9.1_{-5.0}^{+9.6}$ &$1.03_{-0.03}^{+0.21}$ & 37/35 & $31_{-3}^{+4}$ & $46_{-14}^{+20}$\\
\hline
\end{tabular}\\
\vskip 0.3cm
{\it Notes:} {The table gives the median values and their 90\% confidence ranges, while the $\chi^2$ values are given for the best fits. Fixed parameters are marked by `f'. The jet viewing angle, $i_{\rm jet}$, and the difference in the position angles on the sky, $\Delta\lambda$, are derived quantities rather than free parameters. The $b=0.5$ case represents our final results, for which the values of $\langle F_{\rm intr,\nu}\rangle/\langle F_{\nu}\rangle$ refer to the modulated component only.
}
\end{table*}

\begin{figure*}
\centerline{\includegraphics[width=7 cm]{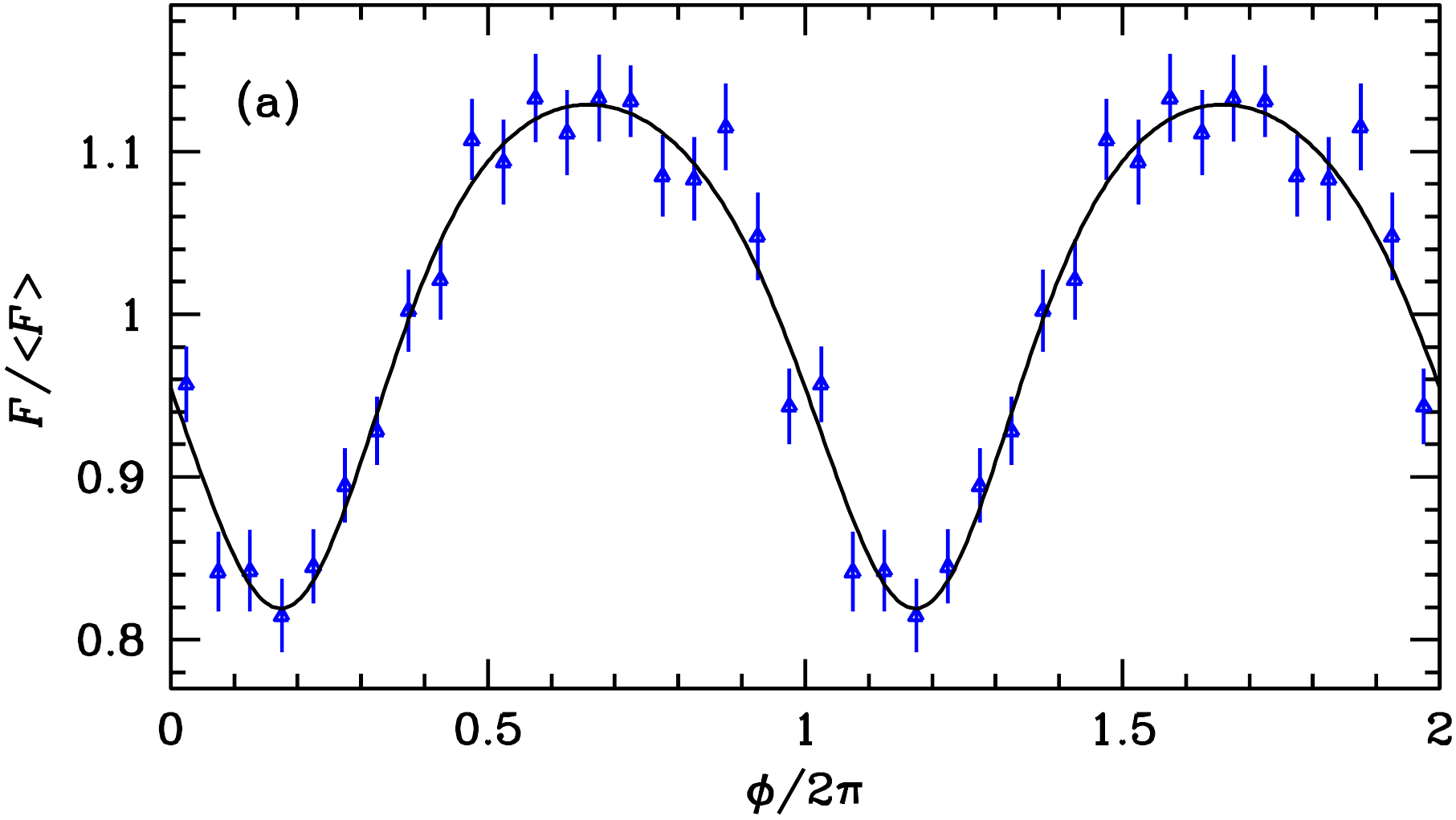} \includegraphics[width=7 cm]{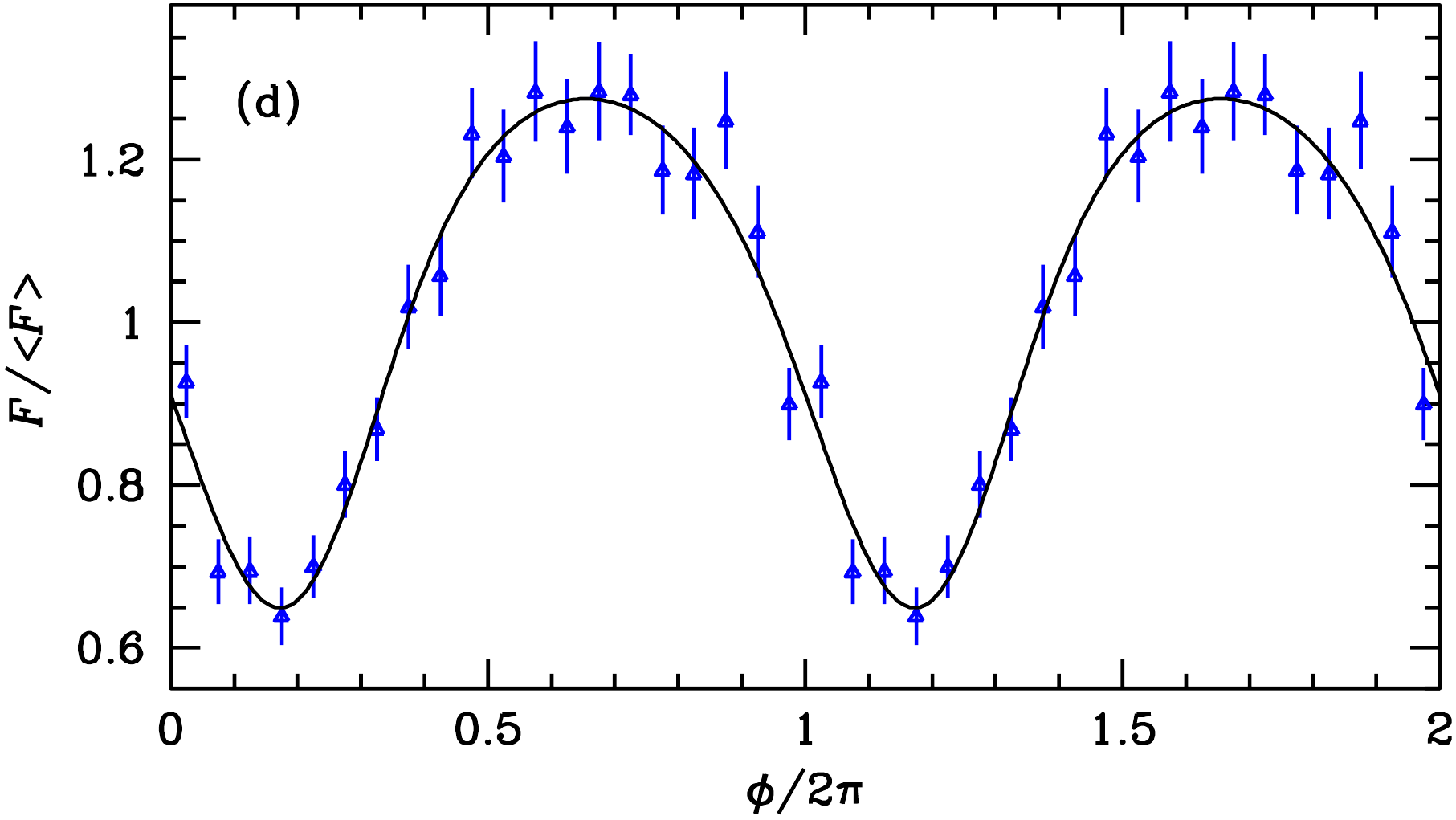}} 
\centerline{\includegraphics[width=7 cm]{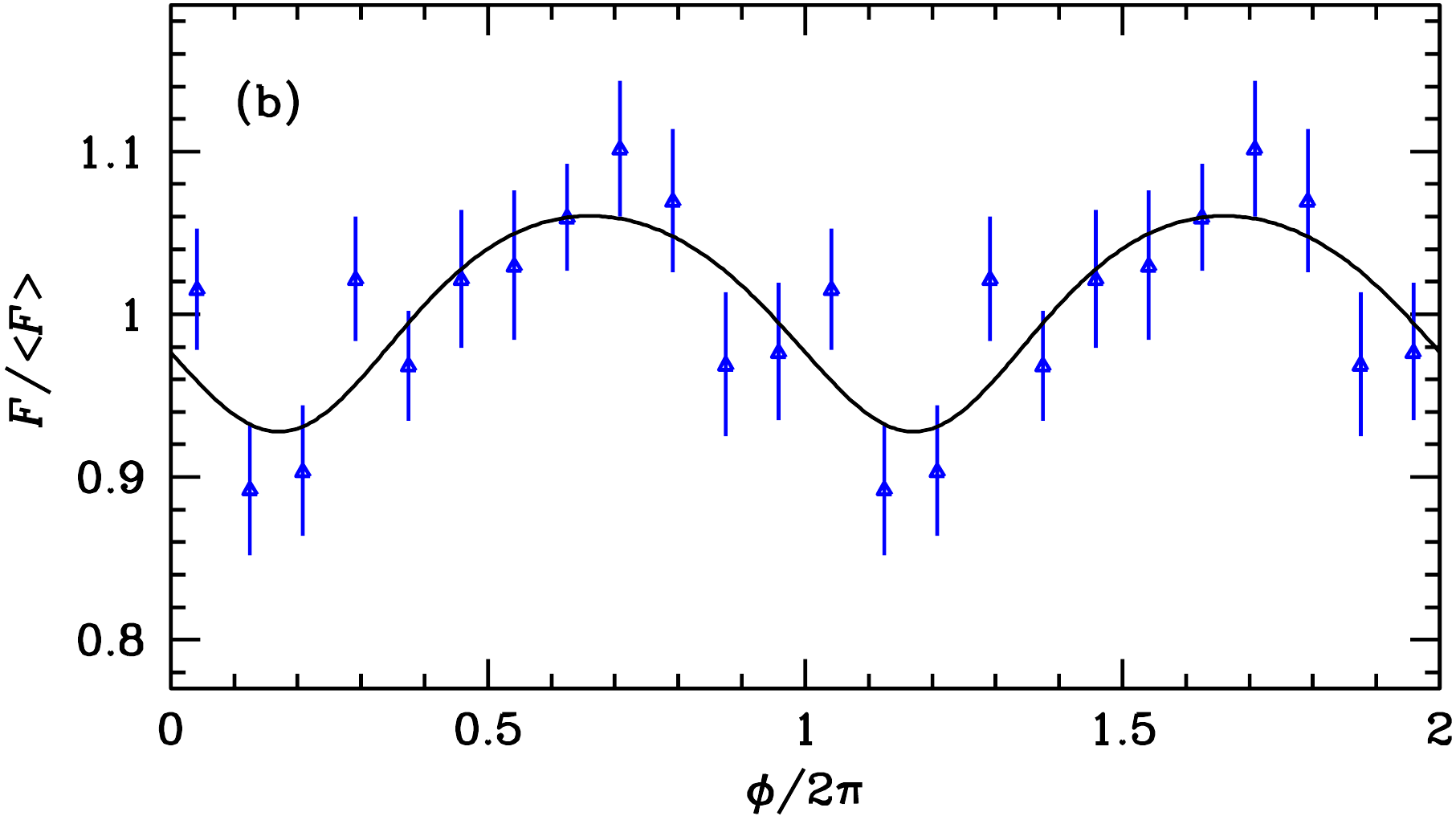} \includegraphics[width=7 cm]{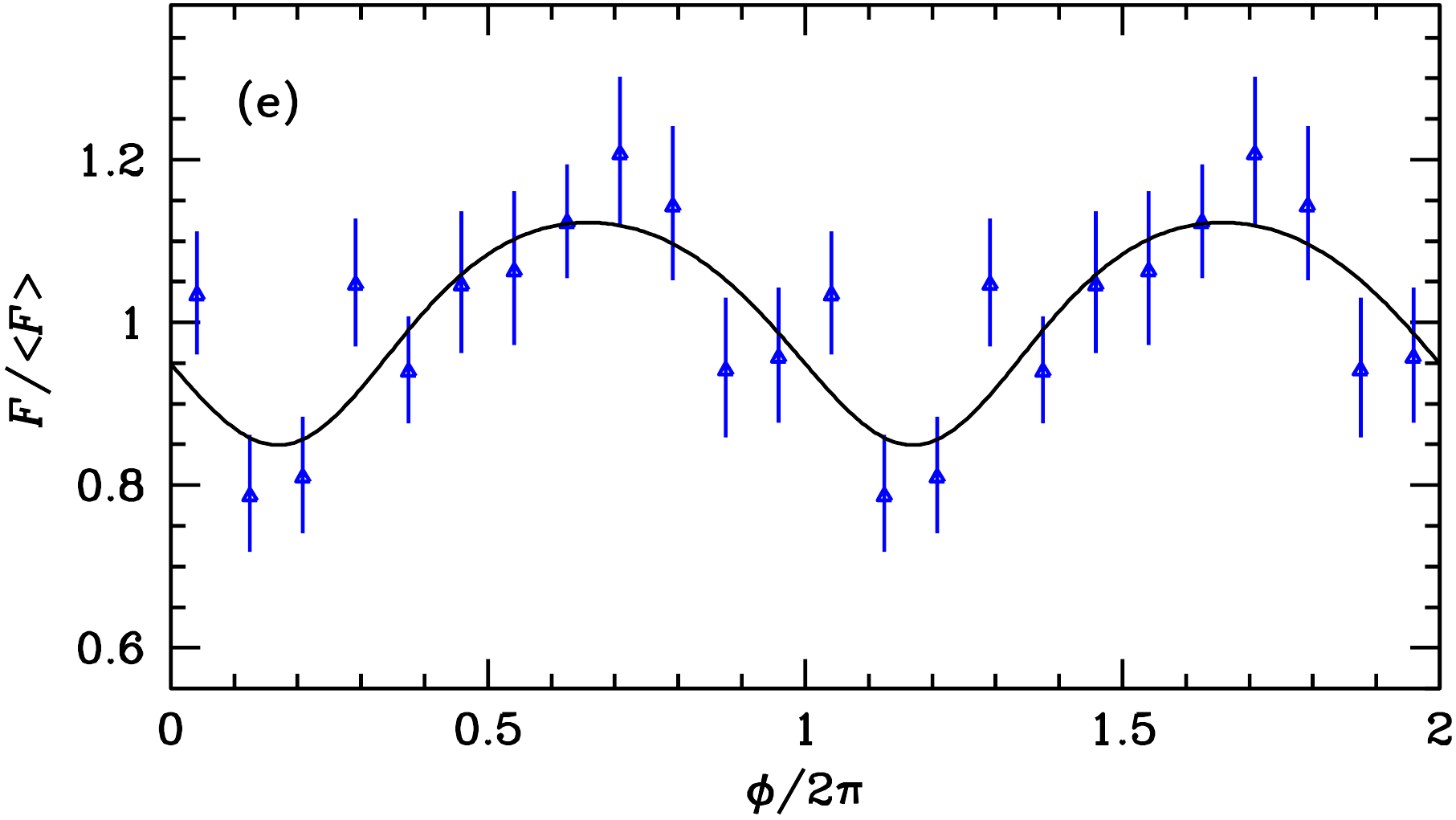}}
\centerline{\includegraphics[width=7 cm]{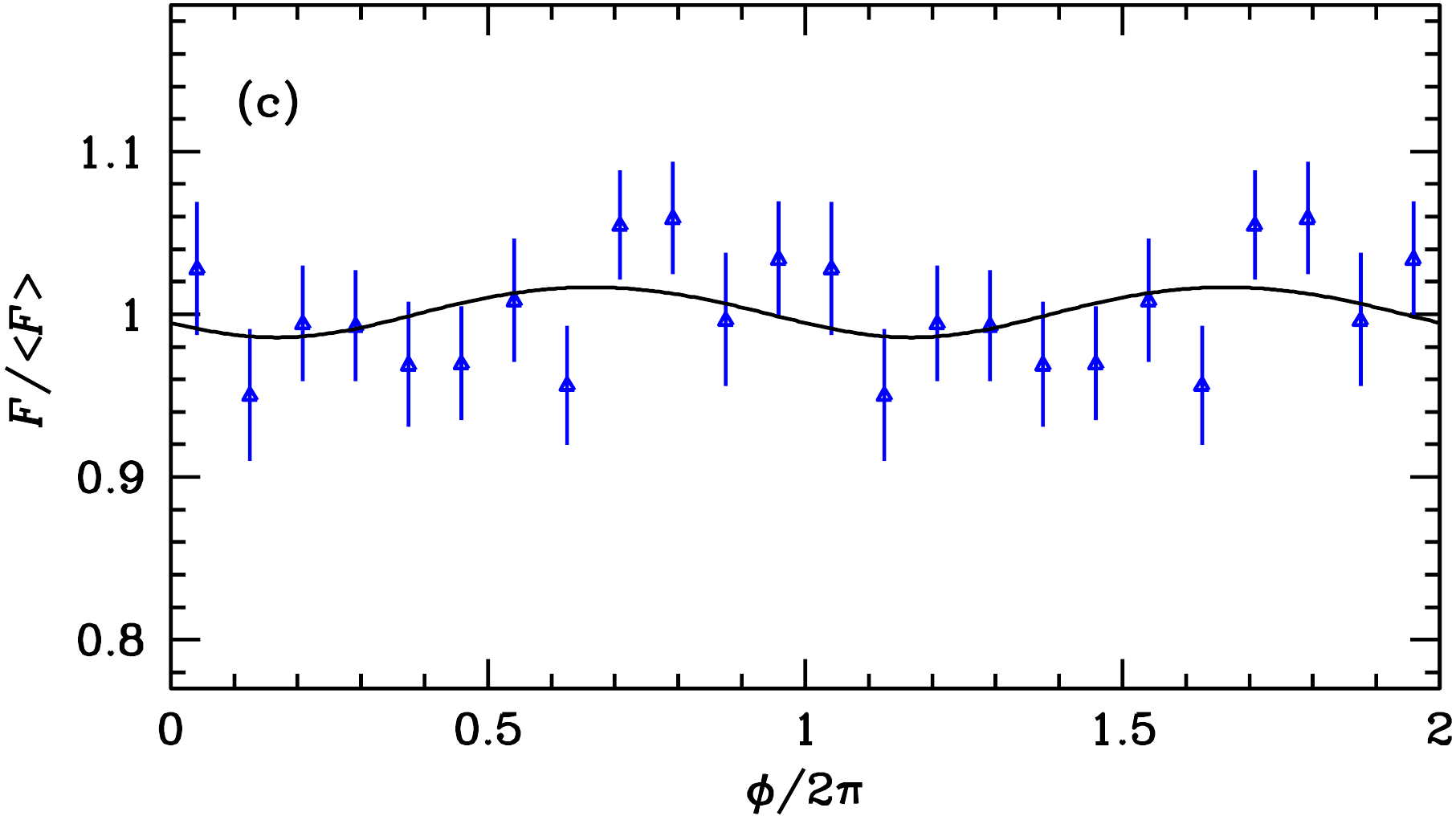} \includegraphics[width=7 cm]{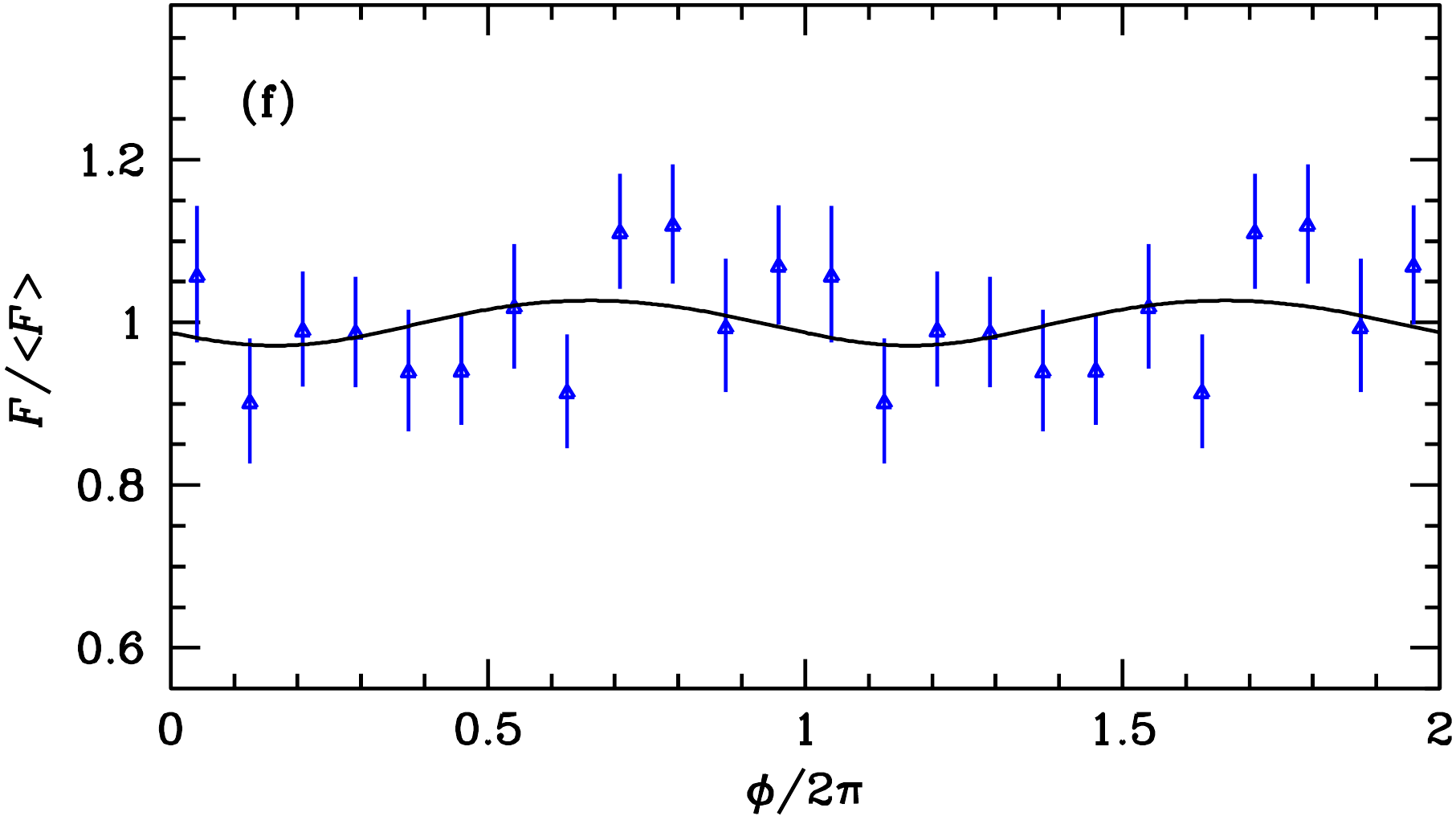}}
\caption{The observed orbital modulations fitted by our model (solid curves). The left (a, b, c) and right (d, e, f) panels assume the 100\% ($b=0$) and 50\% ($b=0.5$) of the total flux is modulated, respectively. (a) 15 GHz, (b) 8.3 GHz, (c) 2.25 GHz,  (d) 15 GHz, (e) 8.3 GHz, (f) 2.25 GHz. For clarity, two cycles are shown. 
}
\label{orb_mod}
\end{figure*}

\begin{figure*}
\centerline{\includegraphics[width=14cm]{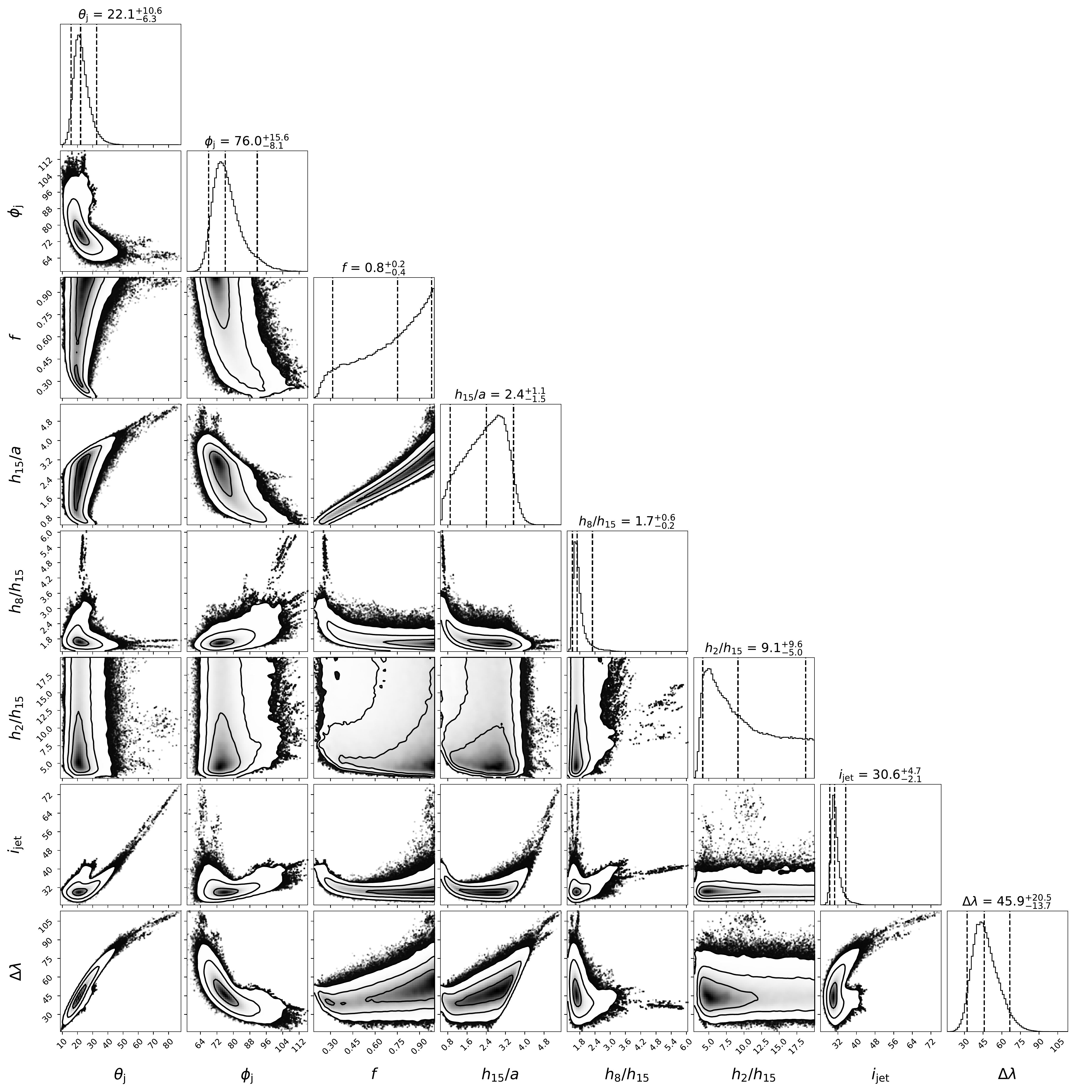}}
\caption{The Markov-chain Monte Carlo results for the case of $b=0.5$. The panels show the histograms of the one-dimensional posterior distributions for the model parameters and the two-parameter correlations. The median results for fitted quantities are shown by the middle vertical dashed lines in the distribution panels. The surrounding vertical dashed lines correspond to the 90\% uncertainty. The parameters obtained are given above the posterior distributions.
}
\label{mcmc}
\end{figure*}

\begin{figure}
\centerline{\includegraphics[width=8cm]{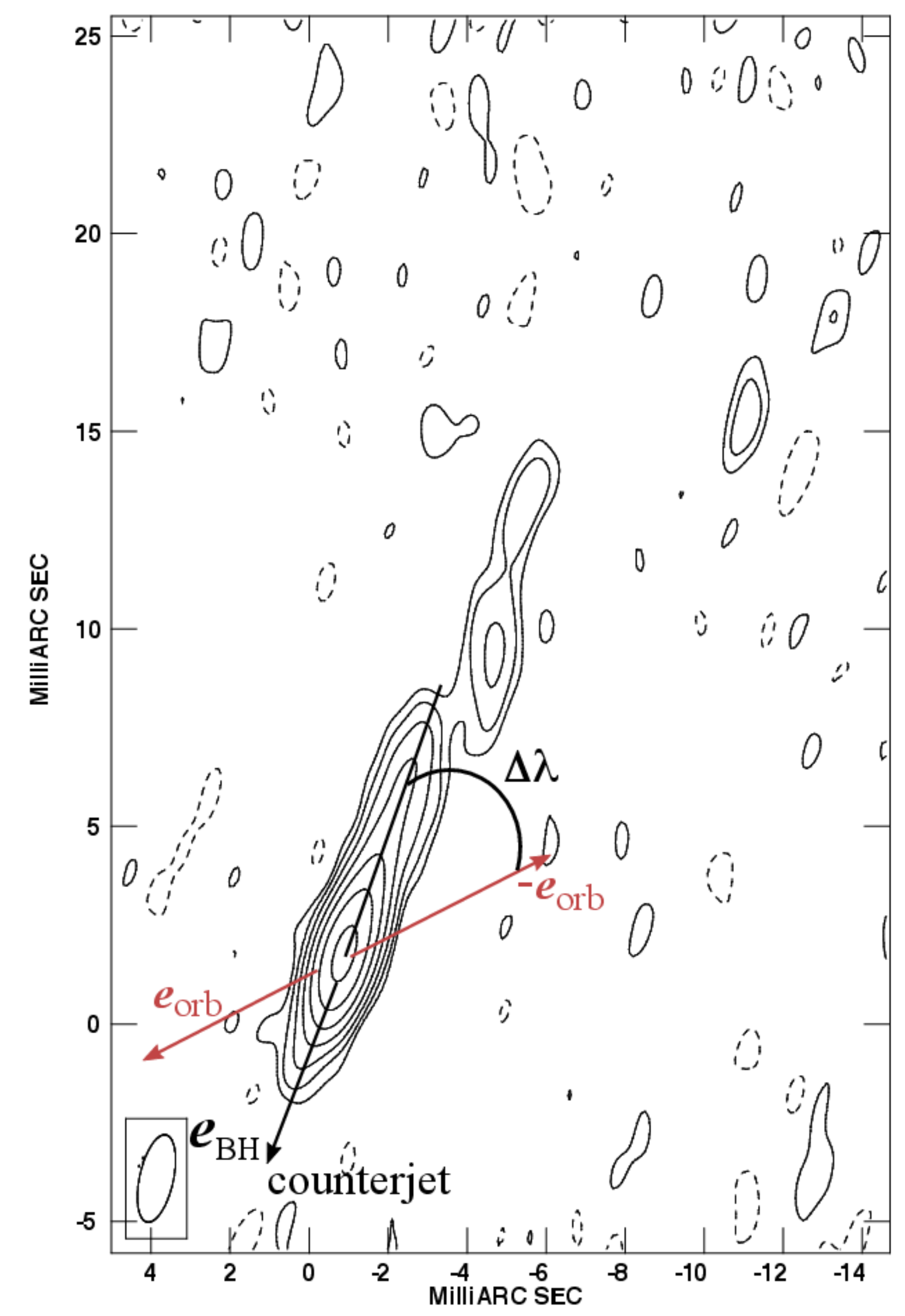}}
\caption{The image of the jet obtained with the VLBA/VLA observation \citep{Stirling01} on 1998 August 10 at 8.4\,GHz. We show the projected position vectors of the orbit and the spin of the BH.}
\label{sky}
\end{figure}

We fit our model to the three folded light curves simultaneously. In order to estimate the uncertainties of the fit, we use both the method based on $\Delta\chi^2$, where $\chi^2$ is the fit statistics \citep{Lampton76} and the Markov Chain Monte Carlo (MCMC) method\footnote{Sanders 2012; \url{https://github.com/jeremysanders/xspec\_emcee}} \citep{Foreman-Mackey13}, both as implemented in {\sc xspec} \citep{Arnaud96}. In the latter, we assume wide normal priors centerd on the best-fit parameters with the widths estimated from the linear error estimates. Both methods give very similar results and we present only those based on the MCMC. The fit results are given in Table~\ref{fits_MCMC}, and the folded light curves are shown in Figure~\ref{orb_mod}. Figure~\ref{mcmc} shows the parameter distributions and corelations for the case of $b=0.5$. The statistical quality of the fits is very good, with the reduced $\chi^2\approx 1$. Table~\ref{fits_MCMC} shows that main difference between the results obtained for $b=0$ and 0.5 is the larger ratios of the intrinsic to the observed fluxes for the modulated component at $b=0.5$. We find a significant BH spin/orbit misalignment of $16\degr$--$33\degr$, and the azimuthal angle of $68\degr$--$92\degr$. These two angles account for the lags seen in the orbital modulations. The emission heights are determined primarily by the modulation amplitudes, and the misalignment angle results jointly from the lags and the heights. The azimuthal angle is close to $90\degr$, i.e., the jet is bent close to the plane of the sky. The attenuation is very modest at each frequency, compatible with the average radio spectrum being approximately a single power law \citep{Fender00}. 

The presence of a misalignment has a very high statistical significance. When we fix the misalignment angle $\Theta_{\rm BH}=0$ (as well as $\phi_{\rm BH}$ at any value, which has then no influence on the fit), we obtain a very high value of $\chi^2$/d.o.f.\ of 402/37 compared to 37/35 for the model with a misalignment. The F-test probability of the fit improvement being by chance \citep{Lampton76} when allowing for a misalignment is $\approx\! 7\times 10^{-19}$. With this model, we also measure the phase lags, i.e., the phases of the minima of the phase-folded fluxes. We keep the assumption of $\Theta_{\rm BH}=0$ but introduce phenomenological phase lags, $\Delta\phi$, with respect to the superior conjunction to the light curves. The obtained values are given in Table~\ref{lags} for $b=0$ and 0.5; the values are almost identical for the two. The hypothesis of no lags has the same statistical significance as that of the lack of a misalignment.

\setlength{\tabcolsep}{3.pt}
\begin{table}
\caption{Estimated phase lags. \label{lags}
}
\centering\begin{tabular}{cccc}
\hline
$b$ & $\Delta\phi_{15}/2\pi$ & $\Delta\phi_{8}/2\pi$ & $\Delta\phi_{2}/2\pi$\\
\hline
0f & $0.17_{-0.01}^{+0.01}$ & $0.17_{-0.07}^{+0.06}$ & $0.35_{-0.15}^{+0.13}$
\vspace {0.1cm}\\
{\bf 0.5}f & $0.17_{-0.01}^{+0.01}$ & $0.17_{-0.07}^{+0.06}$ & $0.35_{-0.16}^{+0.14}$\\
\hline
\end{tabular}\\
\vskip 0.3cm
{\it Notes:} These lags are measured with respect to the model with $\Theta_{\rm BH}=0$ and $\Delta\phi$ added as a phenomenological parameter. Thus, they can slightly differ from the positions of the peaks of the model orbital modulation in Figure \ref{orb_mod}.
\end{table}

Table~\ref{fits_MCMC} gives the jet viewing angle, $i_{\rm jet}$ (Equation~\ref{ibh}). We see that, due to the fitted relative jet orientation with respect to the binary axis and the observer, it is similar to the binary inclination, $i$. On the other hand, we find a relatively large difference between the BH position angle on the sky, $\lambda_{\rm BH}$ (equal to that of the jet $\pm$180$\degr$, equation \ref{delta}) and the implied position angle, $\lambda_{\rm orb}$, of the binary axis, $\Delta\lambda\approx 32\degr$--$66\degr$. The values of $i_{\rm jet}$ and $\Delta\lambda$ depend on all of $i_{\rm orb}$, $\Theta_{\rm BH}$ and $\phi_{\rm BH}$. 

Figure~\ref{sky} shows the image of the jet in Cyg X-1 from a 1998 VLBA/VLA observation \citep{Stirling01}. The position angle of an inner part of the jet is $\lambda_{\rm jet}\approx -17\degr$ (we note it changed to $-26\degr$ in 2016 observations; \mj). We show here the orbital position angle, $\lambda_{\rm orb}$. Due to the obtained large $\Delta\lambda$ (Table~\ref{fits_MCMC}), $\lambda_{\rm orb} \approx 88\degr$--$131\degr$ is very different from the jet position angle.

\section{Discussion}
\label{discussion}

Our model fits very well the data and gives the parameters in full agreement with the standard jet model \citep{BK79}. The fitted height of the 15\,GHz emission  of $2.4^{+1.1}_{-1.5}a$ fully agrees with the estimate of Equation (\ref{hnu}), as well as with more detailed calculations \citep{Zdziarski14c}. Then, the fitted location of the 8.3\,GHz emission agrees (and that of 2.25\,GHz is consistent) with the standard scaling \citep{BK79} of $h\propto \nu^{-1}$.

The misalignment of the BH spin axis with respect to the binary one implies that the spin axis will precess. We use a post-Newtonian estimate \citep{Barker75, Apostolatos94}, which, together with the Kepler law, gives the de Sitter precession period of
\begin{equation}
P_{\rm prec}=\frac{c^2(M_*+M_{\rm BH})^{4/3} P^{5/3}}{(2\pi G)^{2/3}(2+3 M_*/2 M_{\rm BH})M_* M_{\rm BH}},
\label{precession}
\end{equation}
where $G$ is the gravitational constant. For the best-fit parameters of Cyg X-1, $P_{\rm prec}\approx 5.6$\,kyr. While this is a negligible effect for observations of the jet, it has important implications for the origin of the interstellar shell apparently powered by the jet \citep{Gallo05}. In that scenario, the jet lifetime is estimated between 17 and 63\,kyr \citep{Russell07}, i.e., from a few to $\sim\! 10 P_{\rm prec}$. Given our calculated misalignment angle, the jet would affect a larger structure than that observed. The origin of that structure from the jet of Cyg X-1 appears, however, not certain. No analogous structure due to the counterjet has been confirmed \citep{Russell07}, and there is still no definite model of the shell \citep{Sell15}. 

%%% Comparison to other results of orbital/jet orientation

We find the jet viewing angle, $i_{\rm jet}$, is similar to the binary inclination, $i$. This finding does not support the interpretation of the system misalignment along the line of sight, which was put forward as an explanation of the high X-ray polarization \citep{Krawczynski22}. Our results imply that if the inner flow is perpendicular to the BH spin \citep{Bardeen75}, the inclination of that flow is similar to that of the binary, in spite of the jet misalignment. Thus, the strong X-ray polarization has to have another origin, likely coronal outflows \citep{Poutanen23}.

Furthermore, our modelling implies that the jet should be inclined with respect to the binary axis in the plane of the sky; the angle between these two axes is $\Delta\lambda\approx 32\degr$--$66\degr$. This conclusion is surprising in light of the earlier suggestion that the orbital axis, as probed by the position angle of the optical polarization, is well aligned with the position angle of the X-ray polarization and the axis of the radio jet \citep{Krawczynski22, Kravtsov23}. Our findings suggest that the optical polarization should not be related to the binary axis, as previously thought \citep{Kemp78,Kemp79}. Instead, it may arise from scattering in a cocoon surrounding the jet, formed by the jet interaction with the wind of the companion star, or with the interstellar medium \citep{Bicknell97}. In this case, the mean polarization angle would be aligned with the jet. Further thorough analysis of the optical polarization data is needed to verify this suggestion. On the other hand, we have checked that our implied astrometric solution (with the jet misalignment) is approximately consistent with the radio orbital displacements (\mj).

%Discussion of the binary evolution in presence of a misalignment.

The presence of the binary/BH misalignment is in conflict with our current evolutionary scenario for Cyg X-1 (\mj). One possible explanation of this discrepancy is that Cyg X-1 is not a member of the Cyg OB3 association. While the current determinations of their distances, $2.22^{+0.18}_{-0.17}$\,kpc and $1.92\pm 0.31$\,kpc \citep{Rao20}, respectively, are compatible with each other, Cyg X-1 can still be not its member. Projected on the sky, Cyg X-1 is on the edge of the OB3 association, and was considered a field star in a previous clustering analysis \citep{Melnik95}. Alternatively, Cyg X-1 may still be a member of the OB3 association, but the maximum asymmetric natal kick could be higher than the assumed (\mj) 10--20\,km\,s$^{-1}$. In fact, a recent study \citep{Tauris22} has shown that the observed distribution of inspiral spins of merging BHs, showing a strong tail of negative values, is in clear tension with the current theoretical models of binary evolution and supernovae. He showed that the discrepancy can be resolved if some BHs have their spins tossed in a random direction during their formation, and proposed some tentative mechanisms for the tossing. The misalignment in Cyg X-1 can be due to such process.

%%% Alternative interpretations of the phase shifts and whether they can be realised

We then consider alternative interpretations of the phase lags. First, the wind accretion in Cyg X-1 is focused toward the BH with the donor almost filling the Roche lobe (\mj). The wind crossing the L1 point could then be asymmetric, but the maximal absorption is expected at phases before the superior conjunction (due to the Coriolis force; \citealt{Frank87}), contrary to what we see. Furthermore, the X-rays from the system, which originate close to the BH, are orbitally modulated as well, due to bound-free absorption and Compton scattering by the same wind. Their flux minima are observed without any measurable lag \citep{Wen99, Pooley99, Lachowicz06} with respect to the spectroscopically-determined superior conjunction, which shows that any effects of the wind asymmetry on the absorption are negligible. The impact of the wind on the jet can also bend it in the direction along the wind \citep{Bosch-Ramon16}, i.e., with the jet bent outside and precessing at the binary period, which is neither observed in Cyg X-1 nor it would lead to the absorption lags.

Second, the jet trajectory could be helical due to the circular motion of the BH with respect to the center of mass of the binary \citep{SZ07,Bosch-Ramon16}. This could explain the observed radio phase lags, but only if the jet velocity within the distances of $\lesssim\! 30 a$ were nonrelativistic \citep{SZ07}. This disagrees with jet velocity close to the speed of light inferred from the apparent absence of an observed counterjet (according to \citealt{Stirling01}) and the lag of the radio emission with respect to the X-rays \citep{Tetarenko19}. However, this might be still possible if the jet consists of a slow and non-radiating sheath collimating a fast and radiating spine \citep{Ferreira06}, as has been proposed \citep{SZ07}.

Our finding of the jet--orbit misalignment in Cyg X-1 makes it a member of a small group of X-ray binaries with unambiguous observational evidence for misalignments. The others are GRO J1655--40 \citep{Hjellming95,Beer02}, Cyg X-3 \citep{Zdziarski18} and MAXI J1820+070 \citep{Poutanen22}.

\section{Conclusions}

Using an unprecedentedly long light curve of Cyg X-1 at 15\,GHz together with light curves at 8.3 and 2.25 GHz radio frequencies, we obtain strong evidence for a misalignment of its jet with respect to the orbital axis. The misalignment is inferred from the presence of significant delays of the minima of the radio light curves with respect to the times of superior conjunction. The observed strong orbital modulation of the radio flux is due to absorption of the jet emission by the stellar wind of the donor, which depends on the orbital phase. If the jet were aligned with the orbital axis, the maximal absorption -- and the minimum of the radio flux -- would be expected at the phase of the superior conjunction. The observed delays, instead, unambiguously show that the sources of the radio emission are displaced from the symmetry plane of the binary (defined by the semi-major axis and the normal to the binary plane). Thus, the jet at the locations of the radio emission jet is inclined with respect to the binary axis; we find the misalignment angle to be $\approx 16\degr$--$33\degr$. 

However, we find that the jet inclination is approximately the same as the binary inclination, in contrast to the earlier suggestion of a misalignment lying predominantly along the line of sight, motivated by the X-ray polarization studies. Moreover, we find the projection on the sky of the orbital axis is clearly different from that of the radio jet. This finding has implications for the production of the optical polarization. Finally, the presence of the misalignment in Cyg X-1 disagrees at face value with the evolutionary arguments. It implies that either Cyg X-1 is not a member of the Cyg OB3 association or that the kick it received during the BH formation was higher than previously estimated. 

\section*{Acknowledgements}
We thank Richard O'Shaughnessy for a consultation, Rob Fender for permission to use fig.\ 3 of \citet{Stirling01} in our Figure~\ref{sky}, and James Miller-Jones and Arash Bahramian for advice with using their astrometry script. We also thank the referees for valuable comments. We acknowledge the staff who operate and run the AMI-LA telescope at at the Mullard Radio Astronomy Observatory, Lord's Bridge, Cambridge. AMI is supported by the European Research Council under grant ERC-2012-StG-307215 LODESTONE. AAZ and MS acknowledge support from the Polish National Science Center under the grant 2019/35/B/ST9/03944 and from the University of {\L}{\'o}d{\'z} IDUB grant, decision No.\ 59/2021, respectively. Nordita is supported in part by NordForsk. Our work benefited from discussions during Team Meetings in the International Space Science Institute (Bern). 

\bibliography{../../../allbib} 
%\bibliography{allbib} 

\end{document}